\documentclass[a4paper,USenglish,cleveref, autoref, thm-restate]{lipics-v2021}

\hideLIPIcs
\nolinenumbers

\usepackage{proof}
\usepackage{stmaryrd}
\usepackage{xspace}
\usepackage{microtype}
\usepackage{booktabs}
\usepackage[inline,shortlabels]{enumitem}

\usepackage{myisabelle}

\usepackage{tikz}
\usetikzlibrary{shapes.misc}
\definecolor{myBlue}{rgb}{0.17,0.333,0.45}
\definecolor{myGreen}{rgb}{0.3,0.5,0.3}

\definecolor{salmon}{RGB}{250,128,114}
\definecolor{amber}{RGB}{255,191,0}

\definecolor{vscolor}{RGB}{153,76,0}
\definecolor{fscolor}{rgb}{0.3,0.5,0.3}
\definecolor{myBlue}{rgb}{0.17,0.333,0.45}

\newcommand\initial[1]{\textsf{\bfseries #1}}

\newcommand{\relcomp}{\circ}

\newcommand\limplies{\longrightarrow}

\title{A Verified Efficient Implementation of the Weighted Path Order\thanks{The
authors are listed in alphabetical order regardless of individual
contributions or seniority. This research was supported by the Austrian Science Fund (FWF) project I~5943.}}

\titlerunning{A Verified Efficient Implementation of the Weighted Path Order}

\author{René Thiemann}{University of Innsbruck, Austria}{
}{https://orcid.org/0000-0002-0323-8829}{}

\author{Elias Wenninger}{University of Innsbruck, Austria}{}{}{}

\authorrunning{R. Thiemann and E. Wenninger}

\Copyright{René Thiemann and Elias Wenninger}



\supplement{}

\makeatletter
\def\tp@#1#2{\@ifnextchar[{\tp@@{#1}{#2}}{\tp@@@{#1}{#2}}}
\def\tp@@#1#2[#3]#4{#3#1\def\mid{\mathrel{#3|}}#4#3#2}
\def\tp@@@#1#2#3{\bgroup\left#1\def\mid{\;\middle|\;}#3\right#2\egroup}
\def\pa{\tp@()}
\def\tp{\tp@\langle\rangle}
\def\set{\tp@\{\}}
\def\intp{\tp@\llbracket\rrbracket}
\makeatother

\newcommand\mymathchoice[1]{%
    \mathchoice
    {\def\mystyle{\displaystyle}#1}%
    {\def\mystyle{\textstyle}#1}%
    {\def\mystyle{\scriptstyle}#1}%
    {\def\mystyle{\scriptscriptstyle}#1}%
}

\newcommand\SuperImpose[2]{%
    \mymathchoice{%
        \setbox0=\hbox{$\mystyle#1$}
        \setbox1=\hbox{$\mystyle#2$}
        \raisebox{0pt}[\ht0][\dp0]{%
            \ooalign{\hfil\box0\hfil\crcr\hfil\box1\hfil}%
        }%
    }%
}

\newcommand\PGT{>}

\newcommand\GT{\succ}
\newcommand\GE{\succsim}
\newcommand\geopt{%
	\mathrel{\SuperImpose
		\ge
		{\lower.55ex\hbox{{\tiny$($}\phantom{$\ge$}{\tiny$)$}}}
}}
\newcommand\GEopt{%
	\mathrel{\SuperImpose
		\GE
		{\lower.55ex\hbox{{\tiny$($}\phantom{$\GE$}{\tiny$)$}}}
}}
\newcommand\gsopt{%
	\mathrel{\SuperImpose
		\gtrsim
		{\lower.55ex\hbox{{\tiny$($}\phantom{$\sim$}{\tiny$)$}}}
}}
\newcommand\leopt{%
	\mathrel{\SuperImpose
		\le
		{\lower.55ex\hbox{{\tiny$($}\phantom{$\ge$}{\tiny$)$}}}
}}

\newcommand\WPO{\mathsf{WPO}}

\newcommand\TO{\mathsf{WPO}}

\newcommand\lex{\mathsf{lex}}

\newcommand\cF{\mathcal{F}}

\newcommand\cV{\mathcal{V}}

\newcommand\itemstyle[1]{\textcolor{darkgray}{\upshape\sffamily\bfseries\mathversion{bold}#1}}

\newcommand\rEX[1]{Example~\ref{ex:#1}}

\newcommand\isafor{\textsf{Isa\kern-0.15exF\kern-0.15exo\kern-0.15exR}\xspace}
\newcommand\ceta{\textsf{C\kern-0.15exe\kern-0.45exT\kern-0.45exA}\xspace}

\setlist[enumerate,1]{ref   = \arabic*}
\setlist[enumerate,2]{ref   = \theenumi.\alph*}
\setlist[enumerate,3]{ref   = \theenumii.\roman*}
\setlist[enumerate,4]{ref   = \theenumiii.\Alph*}
                      
\begin{document}

\maketitle

\begin{abstract}
The Weighted Path Order of Yamada is a powerful technique for proving
termination. It is also supported by \ceta, a certifier for checking untrusted
termination proofs. To be more precise, \ceta contains a verified function that
computes for two terms whether one of them is larger than the other for a given
WPO, i.e., where all parameters of the WPO have been fixed. The problem of this
verified function is its exponential runtime in the worst case.

Therefore, in this work we develop a polynomial time implementation of WPO that is
based on memoization. It also improves upon an earlier verified implementation
of the Recursive Path Order: the RPO-implementation uses full terms as keys for
the memory, a design which simplified the soundness proofs, but has some runtime
overhead. In this work, keys are just numbers, so that the lookup in the memory
is faster. Although trivial on paper, this change introduces some challenges
for the verification task.
\end{abstract}

\section{Introduction}
\label{sec:intro}

Automatically proving termination of term rewrite systems has been
an active field of research for half a century.
A number of simplification orders~\cite{Dershowitz82,KBO} are classic methods for proving termination,
and these are still integrated in several current termination tools. 
Classical simplification orders are Knuth--Bendix orders (KBO)
and lexicographic and recursive path orders (LPO and RPO).
The weighted path order (WPO) \cite{WPO} was introduced as a
simplification order that unifies and extends classical ones.

Since implementations may be buggy, we are interested in certification, where
automatically generated proofs are checked with verified implementations of
the various orders. Since the verification task is usually time-intensive,
one wants to minimize the number of supported orders. Here, WPO is a perfect candidate 
for integration into a checker, since it covers many simplification orders.  Note that a direct implementation
of WPO as recursive function has an exponential worst case runtime. Therefore, we will develop a non-trivial
implementation that only requires polynomial time. It is based on memoization and its verification
is not immediate.

We perform our formalization using Isabelle/HOL \cite{Isabelle}, based on 
\isafor, the
\initial{Isa}belle \initial{Fo}rmali\-za\-tion of \initial{R}ewriting \cite{isafor},
which also contains soundness proofs of the certifier \ceta.
As a result of this work we improved the execution times of \ceta when it comes to
checking proofs of WPO. We further replaced the existing implementation
of RPO by instantiating the new efficient implementation of WPO,
again leading to faster implementation.
The formalization of the new implementation is available in the archive of formal proofs (AFP)~
\cite{Efficient_Weighted_Path_Order-AFP} and the new implementation is integrated in \ceta version 2.45.

\section{Preliminaries}

We assume familiarity with term rewriting~\cite{BN98},
but briefly
recall notions that are used in the following.
A term built from signature $\cF$ and set $\cV$ of variables
is either $x \in \cV$
or
of form $f(t_1,\dots,t_n)$,
where $f \in \cF$ is $n$-ary and $t_1, \dots, t_n$ are terms.

A reduction pair is a pair $(\succ, \succsim)$ of two relations on terms that
satisfies the following requirements: $\succ$ is
well-founded, $\succsim$ and $\succ$ are compatible (i.e.,
${\succsim\relcomp\succ\relcomp\succsim} \subseteq {\succ}$),
and the relations are closed under certain operations. 

A precedence is a relation $\PGT$ on $\cF$ that is both transitive and well-founded.

We write $\succ^\lex$ and $\succsim^\lex$ for the strict- and non-strict 
  lexicographic extension of $(\succ,\succsim)$. They are defined 
  by a comparison from left to right, e.g., $[s_1,\dots,s_n] \succ^\lex [t_1,\dots,t_n]$ is
  satisfied iff there is some $1 \leq i \leq n$ such that $s_j \succsim t_j$ for all $1 \leq j < i$, and $s_i \succ t_i$.

\section{The Weighted Path Order}

In this section we provide a recursive definition of a variant of WPO. 
We deviate from the original definition by dropping the status function $\pi$ of WPO,
since this feature is not important to illustrate the results of this paper. Actually, the
formalization of WPO \cite{Weighted_Path_Order-AFP} includes several currently known extensions of WPO, e.g.,
quasi-precedences, comparison of variables with least elements, the support of multiset 
comparisons as an alternative to lexicographic comparisons \cite{WPO_WST},
and Refinements (2c) and (2d) of 
WPO \cite[Section 4.2]{WPO}.

Let $\PGT$ be some precedence. 
Let $(\succ,\succsim)$ be some reduction pair such that $\succ$ is transitive, $\succsim$ is a preorder, and $t \succsim s$ for all terms $t$ and
subterms $s$ of $t$.
Then WPO is defined by a strict and non-strict relation on terms ($\GT_{\TO}$ and $\GE_{\TO}$) as follows:
$s \GT_{\TO} t$ iff
\smallskip
	\begin{enumerate}
	\item \label{strict_wpo}
		$s \succ t$, or
	\item \label{weak_wpo}
		$s \succsim t$ and
		\begin{enumerate}
		\item \label{subterm}
			$s = f(s_1,\dots,s_n)$ and $\exists{i \in \{1,\ldots,n\}}.\ s_i \GE_{\TO} t$, or
		\item \label{funfun}
			$s = f(s_1,\dots,s_n)$, $t=g(t_1,\dots,t_m)$ and
			\begin{enumerate}
			\item \label{left_tj} $\forall{j \in \{1,\ldots,m\}}.\ s \GT_{\TO} t_j$ and
			\item \label{precge}
			\begin{enumerate}
			\item
				$f \PGT g$ or
			\item \label{list}
				$f = g$ and $n = m$ and
				$[s_1,\dots,s_n] \GT_{\TO}^{\lex}
				 [t_1,\dots,t_n]$
			\end{enumerate}
			\end{enumerate}
		\end{enumerate}
	\end{enumerate}
\smallskip
The relation $s \GE_{\TO} t$ is defined in the same way,
where $\GT_{\TO}^{\lex}$ in case \itemstyle{\ref{list}} is replaced by
$\GE_{\TO}^{\lex}$, and there is an additional subcase that includes $x \GE_{\TO}^{\lex} x$.
Yamada proved that $(\GT_\TO,\GE_\TO)$ is a reduction pair, and 
can thus be used for termination proofs.

\medskip
The previous verified implementation of WPO just implements the recursive definition of WPO directly.
This can lead to exponentially many recursive calls.

\begin{example}
\label{ex:exp}
We consider a WPO where ${\succ} = \emptyset$ and ${\succsim}$ relates all terms. 
Let us now evaluate $f^{n+1}(x) \GT_{\TO} f^{n+1}(x)$.
We have to consider case~\itemstyle{\ref{subterm}} where we arrive at $f^n(x) \GT_{\TO} f^{n+1}(x)$
and then in case~\itemstyle{\ref{left_tj}} a recursive call to $f^n(x) \GT_{\TO} f^n(x)$ is triggered.
For evaluating $f^{n+1}(x) \GT_{\TO} f^{n+1}(x)$ we further have to apply 
case~\itemstyle{\ref{list}} to arrive at $[f^n(x)] \GT_{\TO}^\lex [f^n(x)]$ which
in turn will again evaluate $f^n(x) \GT_{\TO} f^n(x)$. 

So, evaluation of $f^{n+1}(x) \GT_{\TO} f^{n+1}(x)$ will result in at least two invocations
of $f^n(x) \GT_{\TO} f^n(x)$, which in turn leads to four invocations of
$f^{n-1}(x) \GT_{\TO} f^{n-1}(x)$, then eight invocations of $f^{n-2}(x) \GT_{\TO} f^{n-2}(x)$, etc.
\end{example}

\section{The Efficient Implementation of the Weighted Path Order}

The exponential runtime of WPO as seen in \rEX{exp} is also present in other term orders.
For instance, a memoized implementation of RPO was developed by Nagele to circumvent these problems
\cite{RPO_memo}. There, for an RPO comparison of terms $s$ and $t$ a dictionary was used to lookup and store the results of 
RPO comparisons of subterms.
The keys of the dictionary were pairs of terms $(s_i,t_j)$ where $s_i$ is a subterm of $s$ and $t_j$ is
a subterm of $t$. Consequently, each lookup and store operation on the
dictionary (implemented via red-black trees) requires several term comparisons, which clearly is not optimal when it comes to execution
time. And from the formalization viewpoint it also had its costs, namely the approach required the 
definition of a linear order on terms, which was not automatic at that time.

Similarly, our approach also adds memoization to WPO, but avoiding the problem of term comparisons in dictionary operations.
We use pairs of integer indices as keys instead, which have the advantage of being easy and fast to compare.
The indices are uniquely assigned to all subterms of $s$ and $t$ at the very beginning using an index function.
This index function adds a consecutive integer value to every subterm in the term.

Technically this is done as follows. Terms are represented by a datatype \isa{('f,'v)term}, where \isa{'f} is a type variable for function symbols
and \isa{'v} for variables. Indexed terms are then defined as terms with function symbols and variables adjusted.
\begin{isabelle}
\typesyn ('f,'v)indexed_term = ('f × ('f,'v)term × index, 'v × ('f,'v)term × index)term
\end{isabelle}
\noindent
In this way an indexed term is a term, so one can recurse on its structure. Additionally one has constant time access
to the index of a term (\isa{index :: ('f,'v)indexed_term \To index}) and to the original term 
(\isa{stored :: ('f,'v)indexed_term \To ('f,'v)term}) by accessing the additional informations that are stored in the root symbols,
i.e., variables and function symbols.

The crucial property of the function to create an index (\isa{index_term}) is formulated as:
\begin{isabelle}
\lemma index_term: $\exists$\,ri.\,$\forall$\,t.\,index_term s\,$\unrhd$\,t $ \limplies $ ri\,(index t)\,=\,unindex t $\wedge$ stored t\,=\,unindex t
\end{isabelle} 
\noindent
Here, \isa{unindex} is the function to recursively flatten an indexed term to a term by removing the extra informations in the variables
and function symbols.\footnote{A previous name of the \isa{unindex} function
was \isa{flatten}. This old name is still present in AFP 2022 and \ceta version~2.45, but it will be \isa{unindex} in future versions of the AFP and \ceta.} The existentially quantified
function \isa{ri} (\textbf{r}everse \textbf{i}ndex) is important to show that each index uniquely identifies a subterm.
The equality \isa{unindex (index_term t) = t} is also available.

\medskip

We can now define a memoized implementation of WPO as follows: it takes two indexed terms $s$ and $t$ and a dictionary $d$ as input and then returns
the result $(s \GT_{\WPO} t, s \GE_{\WPO} t)$ in combination with an updated dictionary. Here, \isa{wpo_mem} is the function which tries to
perform a lookup, and if this is not successful, stores the result of the computation of \isa{wpo_main}, which in turn implements the
inference rules of WPO and recurses via
\isa{wpo_mem}.

\begin{isabelle}
\fun wpo_mem \and wpo_main \where
wpo_mem d (s,t) = (\Let i = index s; j = index t \In
  \Case lookup d (i,j) \Of Some result \To (result, d)       (* use memoized result if available *)
    | None \To \Case wpo_main mem (s,t)                         (* otherwise compute result *)
       \Of (result, d') \To (result, update (i,j) result d'))                    (* and memoize it *)
    
wpo_main d (s,t) = (\If stored s $\succ$ stored t \Then ((True, True), d)           (* WPO case 1 *)
  \Else \If stored s $\not\succsim$ stored t \Then ((False, False), d)          (* WPO case 2 not applicable *)
  \Else \Case s \Of Fun f ss $\Rightarrow$ \Case exists_mem ($\lambda$ s$_i$. (s$_i$,t)) wpo_mem snd d ss
     \Of (wpo_2a, d') $\Rightarrow$ \If wpo_2a \Then ((True, True), d')                 (* WPO case 2.a *)
       \Else ...)            
\end{isabelle}

Observe that in the (full) definition of \isa{wpo_main} several higher-order functions have been replaced by higher-order memoized functions, 
e.g., the existential quantification over list elements in WPO case \itemstyle{\ref{subterm}} is
implemented via the memoized version \isa{exists_mem}; similarly there are 
\isa{forall_mem} for case \itemstyle{\ref{left_tj}} and 
\isa{lex_ext_mem} for case \itemstyle{\ref{list}}.

In the memoized implementation of WPO each pair of subterms $s_i \unlhd s$ and $t_j \unlhd t$
is compared at most once, leading to at most $|s| \times |t|$ many invocations of \isa{wpo_main} to compute $s \GT_{\WPO} t$. 
If we assume that
$\succ$ and $\succsim$ can be evaluated in polynomial time, then we get an overall polynomial runtime for the variant of WPO that we
presented here.\footnote{The version of WPO in the AFP includes several improvements. With these improvements, polynomial runtime
is still ensured when fixing a finite signature, but not if arities can grow arbitrarily.}

Now the question is, how we can relate the original implementation \isa{wpo :: ('f,'v)term × ('f,'v)term \To bool × bool}
to its memoized 
implementation \isa{wpo_mem :: (index × index, bool × bool) mapping \To ('f,'v)indexed_term × ('f,'v)indexed_term \To (bool × bool) × (index × index, bool × bool) mapping}.
One prerequisite is to enforce that the dictionary only stores valid results.
Note, however, that in our setting the dictionary stores indices of actual inputs. Therefore, we define a 
valid dictionary for a function \isa{f} as follows, where \isa{rev_ind} is the inverse of an index function,
i.e., it computes from a given index of type \isa{'i} the original value of type \isa{'a} which is then used as input to \isa{f}.

\begin{isabelle}
\definition valid_memory :: ('a \To 'b) \To ('i \To 'a) \To ('i, 'b) mapping \To bool \where
  valid_memory f rev_ind mem = ($\forall$ i b. lookup mem i = Some b $\limplies$ f (rev_ind i) = b)
\end{isabelle}

For proving soundness of \isa{wpo_mem} we require---among others---the precondition \isa{valid_memory wpo ri d} where \isa{ri} is a combination of the
two reverse index functions of $s$ and $t$. Both of these exist by lemma \isa{index_term}.
Note that \isa{ri} does not occur in the implementation.

Before we can prove soundness of \isa{wpo_mem}, we also need to prove soundness of auxiliary memoized functions such as 
\isa{exists_mem}. At this point, let us have a look a bit more closely on how \isa{exists_mem} is actually defined and used.
To this end, consider the type of \isa{exists_mem}.
\begin{isabelle}
\  ('a \To 'b) \To ('m \To 'b \To ('c × 'm)) \To ('c \To bool) \To 'm \To 'a list \To (bool × 'm)
\end{isabelle}

This type clearly deviates from the usual type of a function that checks whether
some list element satisfies a predicate: \isa{('a \To{} bool) \To{} 'a list \To{} bool}.

The second argument to \isa{exists_mem} is a memoized function that, for a given input
of type \isa{'b} and some input memory of type \isa{'m}, computes the result of type \isa{'c} and an updated memory.
Note, however, that the input list takes elements of type \isa{'a} which first need to be converted to inputs for
the memoized function. This is exactly the task of the function that is provided as the first argument of \isa{exists_mem}.
Once we get the result of type \isa{'c} from the memoized implementation, we further need to convert
this into a Boolean to check the predicate, and this is done
by the third argument of \isa{exists_mem}.

In the WPO-example, we need the full flexibility of \isa{exists_mem}.
The memoized function is \isa{wpo_mem} which 
works on pairs of indexed terms and computes pairs of Booleans. Here the first argument to \isa{exists_mem}
is a function that takes some arbitrary subterm $s_i$ from the arguments of $s = f(ss)$ with $ss = [s_1,\dots,s_n]$ and combines
the single indexed term $s_i$ into the pair $(s_i,t)$ such that \isa{wpo_mem} can be invoked on such a pair.
Then the resulting pair $(s_i \GT_{\WPO} t, s_i \GE_{\WPO} t)$ is converted by \isa{snd} into 
$s_i \GE_{\WPO} t$ so that eventually we compute $\exists s_i.\ s_i \GE_{\WPO} t$, i.e., we check 
precisely whether WPO case \itemstyle{\ref{subterm}} is applicable.

For making this reasoning formal, we further need a notion to express that a memoized function (e.g., \isa{wpo_mem})
is an implementation of some function (e.g., \isa{wpo}).

For space reasons we only provide the definition and soundness lemma here, without going into detail.
Note that the handling of the indirection of indices makes the definitions and properties more
complicated than they are in Nagele's formalization without this indirection.

\begin{isabelle}
\definition memoize_fun impl f g rev_ind A =
  ($\forall$ x m y m'. valid_memory f rev_ind m $\limplies$ impl m x = (y,m') $\limplies$ x $\in$ A $\limplies$ 
        y = f (g x) $\wedge$ valid_memory f rev_ind m')
        
\lemma exists_mem: \assumes valid_memory fun rev_ind m
  \and exists_mem f impl h m xs = (b, m') 
  \and memoize_fun impl fun g rev_ind (f ` set xs)
\shows b = ($\exists$ x $\in$ set xs. h (fun (g (f x)))) $\wedge$ valid_memory fun rev_ind m' 
\end{isabelle}

We arrive at the main soundness lemma which shows that \isa{wpo_mem} implements \isa{wpo}. 

\begin{isabelle}
\lemma wpo_mem: \assumes ri = ($\lambda$ (i, j). (rli i, rri j))
  \and \all si. s $\unrhd$ si \implies rli (index si) = unindex si $\wedge$ stored si = flatten si 
  \and \all ti. t $\unrhd$ ti \implies rri (index ti) = unindex ti $\wedge$ stored ti = unindex ti
  \and valid_memory wpo ri d
  \and wpo_mem d (s,t) = (result,d')
\shows result = wpo (unindex s, unindex t) $\wedge$ valid_memory wpo ri d'
\end{isabelle}

We finally define a wrapper that invokes \isa{wpo_mem} with an empty
memory:

\begin{isabelle}
\definition wpo_mem_impl s t = fst (wpo_mem Mapping.empty (index_term s, index_term t))
\end{isabelle}

Soundness of \isa{wpo_mem_impl} is easily proven with the help
of lemmas \isa{index_term} and \isa{wpo_mem}. The \isa{[code]} attribute
tells the code generator of Isabelle to implement \isa{wpo} via the 
defining equations of the memoized implementation.

\begin{isabelle}
\lemma wpo_mem_impl[code]: wpo s t = wpo_mem_impl s t
\end{isabelle}

Based on \isa{wpo_mem} and \isa{wpo_mem_impl} we also obtain \isa{rpo_mem} and \isa{rpo_mem_impl}, a memoized implementation of RPO.
It just instantiates WPO\footnote{Recall
that the formalization contains a variant of WPO that also supports multiset comparisons. In this way RPO
is fully subsumed by WPO.} by choosing the trivial reduction pair, i.e., ${\succ} = \emptyset$
and $\succsim$ compares all terms. In this way, case~\itemstyle{\ref{strict_wpo}} of WPO is never applicable
and the condition $s \succsim t$ in case~\itemstyle{\ref{weak_wpo}} is always satisfied.

\section{Evaluation}

\newcommand\Ff{\mathsf{f}}
\newcommand\Fg{\mathsf{g}}
\newcommand\Fgh{\mathsf{g/h}}
\newcommand\Fh{\mathsf{h}}
\newcommand\Fs{\mathsf{s}}
\newcommand\RR{{\cal R}}
\newcommand\OO{{\cal O}}

The new implementations of WPO and RPO are integrated in \ceta version 2.45 and we experimentally
evaluate it against \ceta version 2.44 which uses a non-memoized WPO and Nagele's implementation
of RPO. For the experiments we use TRSs $\RR_n$ whose termination can be proven by both RPO and WPO.
Eleven of the twelve rules of $\RR_n$ are identical for each $n$, and there is one rule that is parametrized by $n$ as follows.
\begin{equation*}
\Ff(term(n),\Fg(\Fs(y))) \to \Ff(term'(n),\Fs(\Fs(\Fg(y))))
\end{equation*}
Here, both $term(n)$ and $term'(n)$ are terms of the form $(\Fgh)^n(x)$ where each $\Fgh$ is randomly replaced
by either $\Fg$ or $\Fh$. For instance, $term(3)$ might be $\Fg(\Fh(\Fh(x)))$ and $term'(3)$ is $\Fh(\Fh(\Fg(x)))$.
If you download \ceta version 2.45,\footnote{Available at \url{http://cl-informatik.uibk.ac.at/isafor/src/CeTA-2.45.tgz}.} then $\RR_{48}$ is available as \texttt{examples/wpo\_large.proof.xml}, though in a variant
where the $(\Fgh)^{48}(x)$ terms have a non-random structure. 

\begin{figure}
\centering
\includegraphics[width=.5\textwidth]{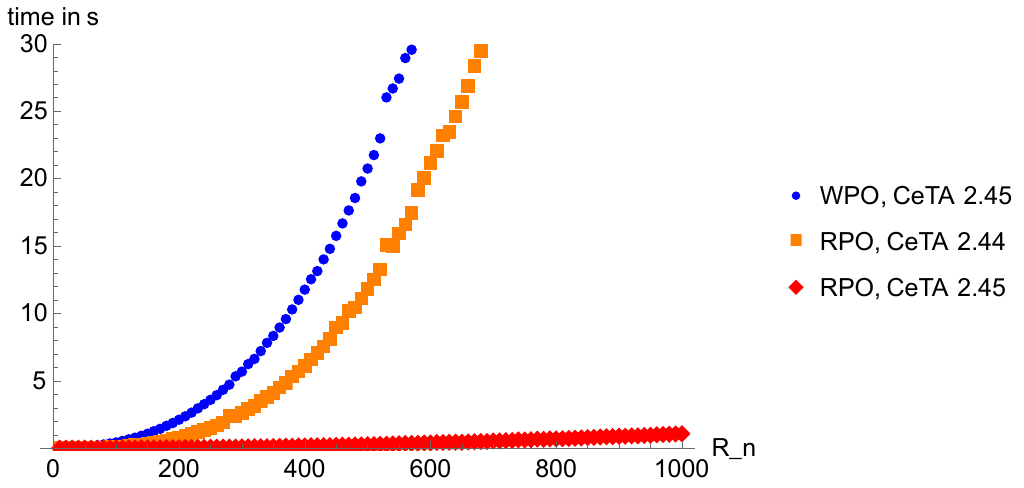}
\caption{Results of running WPO and RPO with \ceta versions 2.44 and 2.45.}
\label{results}
\end{figure}

We tested the implementations on $\RR_{10},\RR_{20},\dots,\RR_{1000}$ and the results are displayed in 
Figure~\ref{results}. All experiments have been conducted using an 3.2 GHz 8-Core Intel Xeon W processor running macOS Ventura 13.4.

\begin{itemize}
\item The RPO implementation in \ceta 2.44 needs cubic time ($\approx 0.09 \cdot n^3$ $\mu$s) to certify the proofs:
  $\OO(n^2)$ many term pairs are compared, and each lookup has a cost of $\OO(n)$. 
\item The WPO implementation in \ceta 2.44 can only solve $\RR_{10}$ and needs 46 seconds. The attempt to certify $\RR_{20}$ was aborted after 10 minutes.
\item The new version of RPO in \ceta 2.45 needs only quadratic time ($\approx 1.10 \cdot n^2$ $\mu$s), since now the lookup costs are negligible.
So, there is a linear speedup compared to \ceta 2.44.
\item The new version of WPO in \ceta 2.45 needs cubic time ($\approx 0.11 \cdot n^3$ $\mu$s), since each comparison in WPO cases~\itemstyle{\ref{strict_wpo}} and
  \itemstyle{\ref{weak_wpo}} requires $\OO(n)$ time for the chosen parameters.  
\end{itemize}

\bibliography{wst_2023}

\end{document}